\documentclass[reprint, superscriptaddress, nofootinbib, pra]{revtex4} 
\usepackage{amssymb, amsmath}
\usepackage{graphicx}
\usepackage{csquotes}
\usepackage{xcolor}
\usepackage[outdir=./]{epstopdf}
\usepackage{verbatim}
\begin{document}
\title{Superdeterministic hidden-variables models II: conspiracy}
\author{Indrajit Sen} \email{isen@clemson.edu}
\affiliation{Department of Physics and Astronomy,\\
Clemson University, Kinard Laboratory,\\
Clemson, SC 29634, USA}
\author{Antony Valentini} \email{antonyv@clemson.edu}
\affiliation{Department of Physics and Astronomy,\\
Clemson University, Kinard Laboratory,\\
Clemson, SC 29634, USA}
\affiliation{Augustus College,\\
14 Augustus Road,\\
London SW19 6LN UK}
\date{\today}

\begin{abstract}
We prove that superdeterministic models of quantum mechanics are conspiratorial in a mathematically well-defined sense, by further development of the ideas presented in a previous article $\mathcal{A}$. We consider a Bell scenario where, in each run and at each wing, the experimenter chooses one of $N$ devices to determine the local measurement setting. We prove, without assuming any features of quantum statistics, that superdeterministic models of this scenario must have a finely-tuned distribution of hidden variables. Specifically, fine-tuning is required so that the measurement statistics depend on the measurement settings but not on the details of how the settings are chosen. We quantify this as the overhead fine-tuning $F$ of the model, and show that $F > 0$ (corresponding to `fine-tuned') for any $N >1$. The notion of fine-tuning assumes that arbitrary (`nonequilibrium') hidden-variables distributions are possible in principle. We also show how to quantify superdeterministic conspiracy without using nonequilibrium. This second approach is based on the fact that superdeterministic correlations can mimic actual signalling. We argue that an analogous situation occurs in equilibrium where, for every run, the devices that the hidden variables are correlated with are coincidentally the same as the devices in fact used. This results in extremely large superdeterministic correlations, which we quantify as a drop of an appropriately defined formal entropy. Nonlocal and retrocausal models turn out to be non-conspiratorial according to both approaches.
\end{abstract}
\maketitle

\section{Introduction}
In a previous article $\mathcal{A}$ \cite{1st}, we gave a broad overview of superdeterministic models and criticisms thereof. We also discussed the properties of these models in nonequilibrium. In particular, we showed that nonequilibrium extensions of superdeterministic models have two striking properties. First, the measurement statistics have a general dependence on the mechanism used to determine the measurement settings. This implies that the equilibrium distribution has to be finely-tuned such that these effects disappear in practice. Second, although these models violate marginal-independence (that is, they violate formal no-signalling), there is no actual signalling. Instead, the local outcomes and distant settings are only statistically correlated due to initial conditions. We argued that this mimicking of a signal using statistical correlations is intuitively conspiratorial. In this paper, we further develop these ideas mathematically to give two separate ways to quantify the conspiratorial character of superdeterministic models.\\ 

It is useful to discuss here the relation between experimental observations, fine tuning, and the nonequilibrium framework. Although all possible distributions are allowed in nonequilibrium, experimental observations are important in the framework. A useful example here would be the astrophysical and cosmological tests of nonequilibrium distributions that have been proposed in pilot-wave theory \cite{valentiniastro, teenv}. The framework, however, considers it unsatisfactory to fine tune the initial hidden-variables distribution to reproduce the experimental observations. A classic example of this is signal locality (marginal independence or formal no-signalling), which has been experimentally confirmed in numerous experiments. The usual explanation of signal locality involves fine tuning of the initial distribution \cite{genon}. The same conclusion was arrived at later by other workers applying causal discovery algorithms to Bell correlations \cite{woodspek}. In this paper, we use the nonequilibrium framework to ask whether superdeterministic models need to be fine tuned so that the measurement statistics depend on the measurement settings but not on the details of how the settings are chosen. \\

The paper is structured as follows. In section \ref{alu1} we define our experimental setup and show that superdeterministic models of our scenario must be fine tuned so that the measurement statistics do not depend on the details of how the measurement settings are chosen. We also show that nonlocal and retrocausal models of our scenario do not have to be similarly fine tuned. In section \ref{alu2}, we quantify the fine tuning required for different kinds of superdeterministic models by introducing a fine tuning parameter $F$. In section \ref{alu3}, we propose a different approach to quantifying the conspiracy which foregoes any use of nonequilibrium distributions. We discuss our results in section \ref{alu4}. 

\section{Superdeterministic conspiracy as fine tuning} \label{alu1}
Consider the standard Bell scenario \citep{bell}, where two spin-1/2 particles prepared in the spin-singlet state are each subjected to one of two local spin measurements, say $M_x$ (corresponding to $\hat{\sigma}_{\hat{x}}$) or $M_z$ (corresponding to $\hat{\sigma}_{\hat{z}}$), in a spacelike separated manner. The local measurement settings at both wings are chosen using a \textit{setting mechanism}, which is defined as any physical system that outputs a measurement setting. A simple example is a die whose faces have been painted either $M_x $ or $M_z$. An experimenter can roll the die for each run and use its output to choose the local measurement setting. Other examples of setting mechanisms can include coins, random number generators, humans choosing measurement settings, and so on. \\

We suppose that the experimenter at each wing has a setup consisting of $N$ different setting mechanisms, which collectively give $N$ outputs for each run. The experimenter then chooses one of the outputs and uses it as the actual measurement setting for that run (how he or she makes this choice has no relevance to our argument, so it can be specified arbitrarily). Nevertheless, the outputs of \textit{all} the $N$ setting mechanisms are recorded for each run at both the wings.\\ 

We label the hidden variables that determine the measurement outcomes at both wings by $\lambda$. The output of the $i^{th}$($j^{th}$) setting mechanism at wing A (B) is labelled by $\alpha_i$ ($\beta_j$), where $i, j \in \{1,2...N\}$ and $\alpha_i, \beta_j \in \{M_x, M_z\}$. We label the experimenters' choice of the outputs in the following manner: if the experimenter at wing A (B) chooses the $i^{th}$ ($j^{th}$) setting mechanism, then we have $\gamma_A = i$ ($\gamma_B = j$). Lastly, we label the actual measurement settings at wing A (B) by $M_A$ ($M_B$), where $M_A, M_B \in \{M_x, M_z\}$.\\

We know that, in a superdeterministic model, $\lambda$ is correlated with the measurement settings. As these models are deterministic, it follows that $\lambda$ is correlated with the physical variables that determine the measurement settings. In our scenario, the measurement settings are functions of the $N$ setting mechanism outputs and of the experimenter's choice for each run (at both wings). Therefore, in general, $\lambda$ will be correlated with $\{\alpha\}$ ($\{\beta\}$) and $\gamma_A$ ($\gamma_B$) at wing A (B), where $\{\alpha\} =\{\alpha_1, \alpha_2,...\alpha_N\}$ and $\{\beta\}=\{\beta_1, \beta_2,... \beta_N\}$. A superdeterministic model of our scenario is schematically illustrated in Fig. 1. Note that, as in $\mathcal{A}$, we have not used the interventionist notion of causation \cite{jearl} to define the causal relationships in the figure, as the model is deterministic. For a deterministic model, the relation between the effect (defined to be in the future) and the cause (defined to be in the past) is that of functional dependence. This definition uses only the concepts of temporal ordering and functional dependence.\\

\begin{figure}
\includegraphics[scale=0.6]{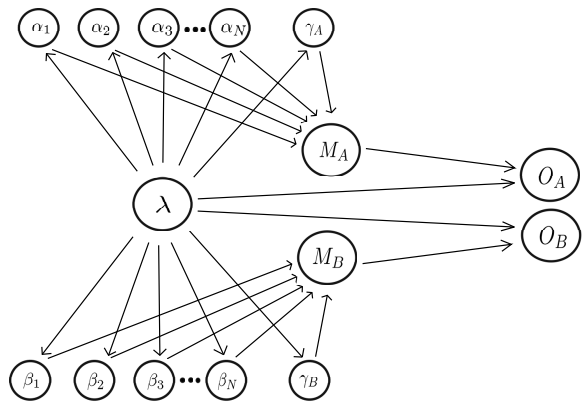}
\caption{Causal order diagram for a superdeterministic model of our scenario. The experimenters choose one of the $N$ setting mechanisms at each wing to select the measurement setting for each run. The setting mechanisms at wing A (B) are depicted by the variables $\alpha$'s ($\beta$'s). The experimenter's choice of setting mechanism at wing A (B) is depicted by $\gamma_A$ ($\gamma_B$). Both type I and type II superdeterministic models are represented by the figure. In a type I model, either $\lambda$ causally influences the variables $\{\alpha\}$, $\gamma_A$ at wing A and $\{\beta\}, \gamma_B$ at wing B, or there are common causes between $\lambda$ and these variables. In the case of common causes, we subdivide $\lambda$ as $\lambda = (\lambda', \mu_A, \mu_B)$, where $\lambda'$ causally influences the outcomes $O_A$ and $O_B$, and $\mu_A$ ($\mu_B$) causally influences both $\lambda'$ and $\{\alpha\}$, $\gamma_A$ ($\{\beta\}, \gamma_B$). A type II model differs from this case only in that $\mu_A$ and $\mu_B$ do not have a causal relationship with $\lambda'$, but are correlated with $\lambda'$ as a direct consequence of the initial conditions.}
\end{figure}

The measurement statistics will be encoded in the distribution $p(O_A, O_B| M_A, M_B)$. For both type I and type II models (see $\mathcal{A}$ and Fig. 1 below), this distribution can be expanded as 
\begin{align}
&p(O_A, O_B| M_A, M_B) = \sum_{\lambda} \sum_{\{\alpha\}, \gamma_A} \sum_{\{\beta\}, \gamma_B} p(O_A, O_B, \lambda, \{\alpha\}, \gamma_A, \{\beta\}, \gamma_B | M_A, M_B)\\
&= \sum_{\lambda} \sum_{\{\alpha\}, \gamma_A} \sum_{\{\beta\}, \gamma_B}p(O_A, O_B| \lambda, M_A, M_B) p(\lambda | M_A, M_B, \{\alpha\}, \gamma_A, \{\beta\}, \gamma_B) p(\{\alpha\}, \gamma_A, \{\beta\}, \gamma_B|M_A, M_B) \label{ow1}
\end{align}
where we have used $ p(O_A, O_B| \lambda,  \{\alpha\}, \gamma_A, \{\beta\}, \gamma_B, M_A, M_B) = p(O_A, O_B| \lambda, M_A, M_B)$ as the measurement outcomes depend only the measurement settings given the hidden variables. We note that the measurement statistics depend, in general, on $\{\alpha\}, \gamma_A$ and $\{\beta\}, \gamma_B$ as the hidden-variables distribution $p(\lambda| M_A, M_B, \{\alpha\}, \gamma_A, \{\beta\}, \gamma_B)$ has a dependence on these variables. For example, suppose the variables are $(\{\alpha\}', \gamma_A', \{\beta\}', \gamma_B')$ for a particular set of runs and $(\{\alpha\}'', \gamma_A'', \{\beta\}'', \gamma_B'')$ for another set. Let the measurement settings for both the sets be the same. That is, $M_A(\{\alpha\}', \gamma_A') = M_A(\{\alpha\}'', \gamma_A'')$ and $M_B(\{\beta\}', \gamma_B') = M_B(\{\beta\}'', \gamma_B'')$. In general, the measurement statistics for both the sets will differ. This can be for two different reasons. First, $\lambda$ may be correlated with different setting mechanisms differently. We know that, when $\gamma_A = i$ ($\gamma_A = j$), $\lambda$ is correlated with $M_A$ due to its correlation with $\alpha_i$ ($\alpha_j$). In general, $\lambda$ may be correlated with $\alpha_i$ and $\alpha_j$ differently. Therefore, if $\gamma_A' = i$ and $\gamma_A'' =j$, ($j \neq i$), then the hidden-variables distribution will be different for both the sets, leading to different measurement statistics in general. Second, $\lambda$ may be non-trivially correlated with the unused setting mechanisms. For example, suppose $\gamma_A' =\gamma_A'' = i$, but $\alpha_j' \neq \alpha_j''$. If $\lambda$ is non-trivially correlated with $\alpha_j$, then the distribution will be different for both the sets, leading to different measurement statistics again in general. Many such sets can be similarly constructed that have the same measurement settings $M_A$ and $M_B$ but different $\{\alpha\}, \gamma_A, \{\beta\}, \gamma_B$ values. The measurement statistics for these sets will be the same if and only if
\begin{align}
&p(O_A, O_B| M_A, M_B, \{\alpha\}', \gamma_A', \{\beta\}', \gamma_B') = p(O_A, O_B| M_A, M_B, \{\alpha\}'', \gamma_A'', \{\beta\}'', \gamma_B'') \label{ii}
\end{align}
where $M_A(\{\alpha\}', \gamma_A') = M_A(\{\alpha\}'', \gamma_A'')$ and $M_B(\{\beta\}', \gamma_B') = M_B(\{\beta\}'', \gamma_B'')$. Equation (\ref{ii}) formalises the condition that the measurement statistics depend only on the measurement settings. Expanding both distributions over $\lambda$, we have
\begin{align}
\sum_{\lambda} p(O_A, O_B| \lambda, M_A, M_B) p(\lambda| M_A, M_B, \{\alpha\}', \gamma_A', \{\beta\}', \gamma_B') = \sum_{\lambda} p(O_A, O_B| \lambda, M_A, M_B) p(\lambda| M_A, M_B, \{\alpha\}'', \gamma_A'', \{\beta\}'', \gamma_B'') \label{i2}
\end{align}
It is clear that condition (\ref{i2}) will not be satisfied in general for an arbitrary hidden-variables distribution $p(\lambda| M_A, M_B, \{\alpha\}, \gamma_A, \{\beta\}, \gamma_B)$. Therefore, we conclude that superdeterministic models of our scenario must be fine tuned. Note that this fine tuning is not needed to reproduce any particular feature of the quantum predictions, such as marginal independence (formal no signalling), non contextuality, and so on. It is only needed to satisfy equation (\ref{ii}). \\

On the other hand, one can show that retrocausal and nonlocal models of our scenario do not need this finetuning\footnote{All hidden-variables models need to be fine-tuned in order to reproduce marginal independence (formal no signalling) \cite{genon, woodspek}, unless a dynamical explanation of finetuning is given \cite{valentinI, royalvale, teenv}. The fine tuning we are discussing here refers only to that required to satisfy condition (\ref{ii}).}. Let us first consider retrocausal models of our scenario. The causal order diagram for such a model is shown in Fig. 2 a). Note that we have used the interventionist notion of causation \cite{jearl} to define causal relationships in Fig. 2 a) as retrocausal models are not deterministic (given the past conditions alone). The measurement setting causally affects $\lambda$ backwards in time in these models. The measurement setting, in turn, is causally determined by the setting mechanisms and the the experimenters' choices forwards in time. In this way, $\lambda$ is correlated with the variables $\{\alpha\}, \gamma_A$ at wing A and $\{\beta\}, \gamma_B$ at wing B. Intuitively, we expect that only the measurement setting, not the exact manner in which the measurement setting is chosen, will be important. This is because the setting of the measuring apparatus, regardless of how it is selected, causally affects $\lambda$ (backwards in time). Therefore, the correlation between $\lambda$ and the measurement settings does not depend on how the settings are chosen, that is, $p(\lambda|M_B, \{\alpha\}, \gamma_A, \{\beta\}, \gamma_B) = p(\lambda|M_B)$ for the retrocausal model in Fig. 2 a). We can understand this more explicitly by considering the joint distribution of the model parameters
\begin{align}
p(O_A, O_B, \lambda, M_A, M_B, \{\alpha\}, \gamma_A, \{\beta\}, \gamma_B) = &p(O_A|\lambda, M_A) p(O_B|\lambda, M_B) p(\lambda| M_B) p(M_A|\{\alpha\}, \gamma_A) \times \nonumber\\
&p(M_B| \{\beta\}, \gamma_B) p(\{\alpha\}) p(\gamma_A) p(\{\beta\}) p(\gamma_B)
\end{align}
We can sum over the variables $\{\alpha\}, \gamma_A$ at wing A and $\{\beta\}, \gamma_B$ at wing B to yield
\begin{align}
\sum_{\{\alpha\}, \gamma_A} \sum_{\{\beta\}, \gamma_B} p(O_A, O_B, \lambda, M_A, M_B, \{\alpha\}, \gamma_A, \{\beta\}, \gamma_B) = p(O_A|\lambda, M_A) p(O_B|\lambda, M_B) p(\lambda|M_B) p(M_A) p(M_B)
\end{align}
This means that we can always build a reduced model where only the parameters $p(O_A|\lambda, M_A)$, $p(O_B|\lambda, M_B)$, $p(\lambda|M_B)$ and $p(M_A)$, $p(M_B)$ are defined. The variables $\{\alpha\}, \gamma_A, \{\beta\}, \gamma_B$ do not need to be explicitly included in the model. Therefore, the model does not have to be fine tuned with respect to these variables and condition (\ref{ii}) is automatically satisfied. \\

Next consider a nonlocal model of our scenario. The causal order diagram will be as shown in Fig. 2 b). The joint distribution of the model parameters is 
\begin{align}
p(O_A, O_B, \lambda, M_A, M_B, \{\alpha\}, \gamma_A, \{\beta\}, \gamma_B) = &p(O_A|\lambda, M_A, M_B) p(O_B|\lambda, M_B) p(\lambda) p(M_A|\{\alpha\}, \gamma_A) \times \nonumber\\
&p(M_B| \{\beta\}, \gamma_B) p(\{\alpha\}) p(\gamma_A) p(\{\beta\}) p(\gamma_B)
\end{align}
Similarly to the retrocausal model, we can sum over the variables $\{\alpha\}, \gamma_A$ at wing A and $\{\beta\}, \gamma_B$ at wing B to yield
\begin{align}
\sum_{\{\alpha\}, \gamma_A} \sum_{\{\beta\}, \gamma_B} p(O_A, O_B, \lambda, M_A, M_B, \{\alpha\}, \gamma_A, \{\beta\}, \gamma_B) = p(O_A|\lambda, M_A, M_B) p(O_B|\lambda, M_B) p(\lambda) p(M_A) p(M_B)
\end{align}
Therefore, the only necessary model parameters are $p(O_A|\lambda, M_A, M_B)$, $p(O_B|\lambda, M_B)$, $p(\lambda)$ and $p(M_A)$, $p(M_B)$. A nonlocal model of our scenario, therefore, does not have to be fine tuned with respect to the variables $\{\alpha\}, \gamma_A, \{\beta\}, \gamma_B$ in order to satisfy condition (\ref{ii}). Unlike for retrocausal and nonlocal models, the variables $\{\alpha\}, \{\beta\}, \gamma_A,$ and $\gamma_B$ cannot be summed over similarly for superdeterministic models of our scenario. The causal order diagram for a superdeterministic model of our scenario (see Fig. 1) implies that $p(\lambda|M_A, M_B, \{\alpha\}, \gamma_A, \{\beta\}, \gamma_B) = p(\lambda|\{\alpha\}, \gamma_A, \{\beta\}, \gamma_B)$. \\

The measurement statistics for superdeterministic models depend crucially on the correlation between $\lambda$ and the variables $\{\alpha\}, \{\beta\}, \gamma_A,$ and $\gamma_B$ (given $M_A$ and $M_B$). Therefore, fine tuning is required so that the measurement statistics depend on the measurement settings but not on the details of how the settings are chosen. Note that we have not imposed any features of quantum statistics, that is, any features of the Born rule, to argue that superdeterministic models are fine tuned. In principle, we can impose the additional requirement that the superdeterministic model reproduces the Born rule. This will require an additional fine tuning. However, this is not relevant to the question we are asking in this paper: whether superdeterministic models require fine tuning so that the measurement statistics depend on the measurement settings but not on the details of how the settings are chosen. Our results show that superdeterministic models require fine tuning regardless of whether or not they reproduce quantum statistics, in sharp contrast with retrocausal or nonlocal models. Note further that we have not imposed the condition of locality either.\\

\begin{figure}
\includegraphics[scale=0.6]{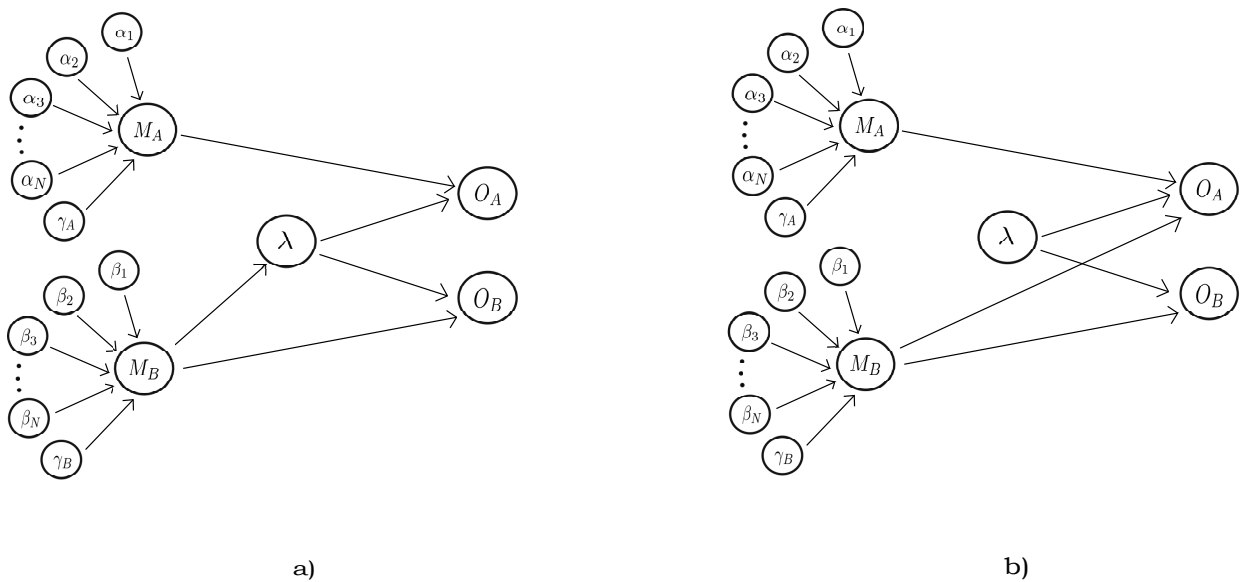}
\caption{Schematic illustrations of retrocausal and nonlocal models of our scenario. There are $N$ setting mechanisms at each wing. For each run, the experimenters choose one of the setting mechanisms to select the measurement setting for that run. Part $a)$ illustrates a retrocausal model of our scenario. The measurement setting $M_B$ causally affects the hidden variables $\lambda$ backwards in time. This leads to a correlation between the variables $\{\beta\}, \gamma_B$ and $\lambda$. Part $b)$ illustrates a nonlocal model of our scenario. The measurement setting $M_B$ nonlocally affects the outcome $O_A$. In both cases, it is possible to sum over the $\{\alpha\}, \gamma_A$ variables at wing A and $\{\beta\}, \gamma_B$ variables at wing B to build a reduced model without these variables.}
\end{figure}

\section{Quantification of fine tuning} \label{alu2}
In this section, we quantify the degree of fine tuning required for superdeterministic models of our scenario to satisfy condition (\ref{ii}). We first assume that 
\begin{align}
p(\lambda| M_A, M_B, \{\alpha\}, \{\beta\}, \gamma_A, \gamma_B) = l/L \textbf{ }\textbf{ }\forall\textbf{ } \lambda \label{i20}
\end{align}
for some $l \in \{0,1,2...L\}$, where $L$ is a very large positive integer. Although we are assuming that $\lambda$ is a discrete variable, our argument can be applied to discrete approximations of any continuous distribution, where the approximation can be made arbitrarily close by increasing $L$. Let the number of possible values of $\lambda$ be $\Lambda$.\\

We can represent the hidden-variable distribution by using the probability simplex $\big (p \in \mathcal{R}^{\Lambda}_+ | \sum_{\lambda} p(\lambda|M_A, M_B, \{\alpha\}, \{\beta\}, \gamma_A, \gamma_B) =1, p(\lambda|M_A, M_B, \{\alpha\}, \{\beta\}, \gamma_A, \gamma_B) \geq 0 \big )$. In the simplex notation, a particular configuration of the distribution is represented by a single point $p$ on a $(\Lambda -1)$ dimensional object. By a configuration, we mean a particular value of the distribution for all $\lambda$'s. For example, for $\Lambda = 3$, the different configurations of the distribution can be represented by points on a plane (see Fig. 3). Each point on the plane specifies a particular set of values $\{p(\lambda_1|M_A, M_B, \{\alpha\}, \{\beta\}, \gamma_A, \gamma_B), p(\lambda_2|M_A, M_B, \{\alpha\}, \{\beta\}, \gamma_A, \gamma_B), p(\lambda_3|M_A, M_B, \{\alpha\}, \{\beta\}, \gamma_A, \gamma_B)\}$ for the distribution $p(\lambda|M_A, M_B, \{\alpha\}, \{\beta\}, \gamma_A, \gamma_B)$. Let the number of points on the simplex that satisfy (\ref{i20}) for all $\lambda$ be labelled by $V(\Lambda, L)$. In nonequilibrium, there are $V(\Lambda, L)$ possible configurations of each distribution. Note that $V(\Lambda, L)$ scales with $\Lambda$ as $V(\Lambda, L) = (L + \Lambda -1)!/ L! (\Lambda-1)!$.\\

Let $\Omega$ be the total number of independent distributions in the model, including nonequilibrium distributions. For example, consider $\Omega$ for our scenario. We know that there are $N$ possible $\gamma_A$ and $\gamma_B$ values. Each $\alpha$ or $\beta$ can take one of 2 values ($M_x$ or $M_z$). The number of possible $\{\alpha\}$ or $\{\beta\}$ values, therefore, is $2^{N}$. In total, there will be $N^2 2^{2N}$ independent distributions $p(\lambda| M_A, M_B, \{\alpha\}, \{\beta\}, \gamma_A, \gamma_B)$. Thus, $\Omega = N^2 2^{2N}$ for our scenario. Each distribution can take one of $V(\Lambda, L)$ configurations (points on the simplex). The $\Omega$ distributions can have $V(\Lambda, L)^{\Omega}$ different configurations in total. Let the model impose certain constraints on the configurations to satisfy condition (\ref{ii}). Suppose, as a result of these constraints, the total number of configurations is reduced to $N_f$. Then we define the quantity
\begin{align}
F =  1 - \frac{N_f}{V(\Lambda, L)^{\Omega}} \label{main}
\end{align}
as the overhead fine-tuning of the hidden-variables model. If $F=0$ we call the model \textit{completely general}, as the total number of configurations after applying the constraints is the same as before. On the other hand, if $F=1$ we call the model \textit{completely fine-tuned}, as there are no configurations left after imposing the constraints. Any value of $F$ outside $[0,1]$ reflects inconsistency in the model. We now use this definition to quantify the fine tuning of superdeterministic models of our scenario.\\

\begin{figure}
\includegraphics[scale=0.7]{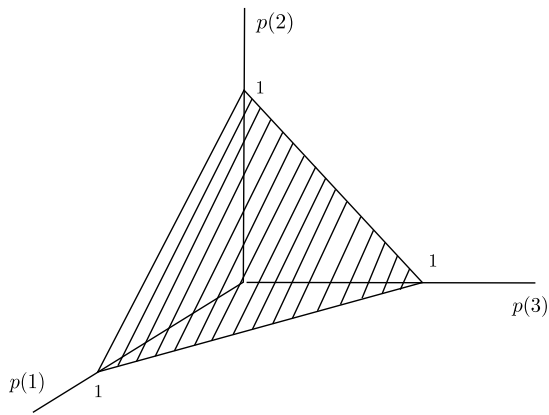}
\caption{Schematic illustration of the probability simplex for $\Lambda =3$. Each point on the shaded plane represents a particular distribution $\{p(1), p(2), p(3)\}$. In general, a hidden-variables distribution can be represented by a point on a $(\Lambda -1)$ dimensional space in this fashion for any $\Lambda$.}
\end{figure}

We first consider the kind of superdeterministic model usually proposed in the literature, where there is single distribution $p(\lambda| M_A, M_B)$. One can obtain such a model by imposing the following constraints on the general distribution $p(\lambda| M_A, M_B, \{\alpha\}, \{\beta\}, \gamma_A, \gamma_B)$:
\begin{align}
p(\lambda| M_A, M_B, \{\alpha\}, \{\beta\}, \gamma_A=i, \gamma_B=j) &= p(\lambda| M_A, M_B, \alpha_{i}, \beta_{j}, \gamma_A=i, \gamma_B=j) \label{m}\\
p(\lambda| M_A, M_B, \alpha_{i}, \beta_{j}, \gamma_A=i, \gamma_B=j) &= p(\lambda| M_A, M_B) \label{n}
\end{align}
for all $i, j$. Note that $M_A = M_A(\{\alpha\}, \gamma_A)$ and $M_B = M_B(\{\beta\}, \gamma_B)$. Constraint (\ref{m}) implies that $\lambda$ is correlated with the output of only the setting mechanism actually used for all runs. Constraint (\ref{n}) implies that $\lambda$ is correlated with the output of each setting mechanism in the same way. Note that $p(\lambda| M_A, M_B)$ need not be the equilibrium distribution, as we have not imposed the condition that the model reproduces the predictions of quantum mechanics. It is clear that a superdeterministic model that satisfies equations (\ref{m}) and (\ref{n}) will satisfy equation (\ref{i2}), and thereby condition (\ref{ii}). Let us find out the value of $F$ for such a model.\\

The model (after being constrained) has only four independent distributions $p(\lambda| M_A, M_B)$, as we assume that $M_A$ and $M_B$ can each take one of only two values: $M_x$ or $M_z$. Each of the four distributions can occupy any of the $V(\Lambda, L)$ points on the simplex, as (\ref{i2}) is satisfied automatically due to (\ref{m}) and (\ref{n}). Therefore, $N_f = V(\Lambda, L)^4$. The fine-tuning parameter $F$ will be
\begin{align}
F &= 1 - \frac{V(\Lambda, L)^4}{V(\Lambda, L)^{\Omega}}\\
&= 1 - V(\Lambda, L)^{4 -N^2 2^{2N}},
\end{align}
as we know that $\Omega = N^2 2^{2N}$. From the `excess-baggage' theorem \citep{hardycess}, we know that $\Lambda \to \infty$. For $N= 1$, $F= 0$ (there is no fine tuning). However, for any $N > 1$, $F = 1$ in the limit $V(\Lambda, L) \to \infty$ (we know that as $\Lambda \to \infty$, $ V(\Lambda, L) \to \infty$). Therefore, the model is arbitrarily close to complete fine tuning for any $N > 1$. This means that imposing the constraints (\ref{m}) and (\ref{n}) to reproduce the features $a)$ and $b)$ is a bad strategy from a fine tuning perspective. \\

Let us now consider more general superdeterministic models. We know that equation (\ref{i2}) is the minimum constraint that needs to be satisfied in order to satisfy (\ref{ii}). Let us, therefore, determine $F$ for a superdeterministic model of our scenario with (\ref{i2}) as the only constraint. This will be the minimum fine tuning required for a general superdeterministic model of our scenario. To begin, we note that the total number of independent distributions does not change as a result of (\ref{i2}). Out of these $\Omega$ distributions, there are $\Omega/4$ distributions corresponding to a particular value of $(M_A, M_B)$. That is, there are $\Omega/4$ distributions corresponding to each of $(M_x, M_x)$, $(M_x, M_z)$, $(M_z, M_x)$ and $(M_z, M_z)$. We need to ensure that equation (\ref{i2}) is satisfied for each value of $(M_A, M_B)$. Consider the $\Omega/4$ distributions that correspond to a particular value of $(M_A, M_B)$. Consider a single distribution out of these possibilities. Let this distribution be unconstrained, so that the number of possible configurations of this distribution is $V(\Lambda, L)$. Let us label the $j^{th}$ configuration of the unconstrained distribution by $p^j(\lambda| M_A, M_B, \{\alpha\}^u, \{\beta\}^u, \gamma_A^u, \gamma_B^u)$, where $j \in \{1, 2, 3....V(\Lambda, L)\}$. Consider another (constrained) distribution $p(\lambda| M_A, M_B, \{\alpha\}^c, \{\beta\}^c, \gamma_A^c, \gamma_B^c)$ out of the remaining $\Omega/4 - 1$ possible distributions that correspond to $(M_A, M_B)$. Equation (\ref{i2}) then implies
\begin{align}
\sum_{\lambda} p(O_A, O_B| \lambda, M_A, M_B) p^j(\lambda| M_A, M_B, \{\alpha\}^u, \gamma_A^u, \{\beta\}^u, \gamma_B^u) = \sum_{\lambda} p(O_A, O_B| \lambda, M_A, M_B) p(\lambda| M_A, M_B, \{\alpha\}^c, \gamma_A^c, \{\beta\}^c, \gamma_B^c) \label{i2'}
\end{align}
Clearly, the remaining $\Omega/4 - 1$ distributions will have to be constrained in order to satisfy (\ref{i2'}). Let the number of configurations of each constrained distribution (that satisfies (\ref{i2'}) for a given $j^{th}$ configuration of the unconstrained distribution) be $v^j_{AB}(\Lambda, L)$, where $0 < v^j_{AB}(\Lambda, L)/V(\Lambda, L) < 1$. The total number of configurations of the distributions corresponding to $(M_A, M_B)$ will then be $\sum_{j=1}^{V(\Lambda, L)} v^j_{AB}(\Lambda, L)^{\frac{\Omega}{4} -1}$.  The fine tuning parameter $F$ will be
\begin{align}
F = 1 -  \frac{\prod_{A, B} \sum_{j=1}^{V(\Lambda, L)} v^j_{AB}(\Lambda, L)^{\frac{\Omega}{4} -1}}{V(\Lambda, L)^{\Omega}}
\end{align}
We know that $v^j_{AB}(\Lambda, L)/V(\Lambda, L) < 1$ $\forall \textbf{ } j, A, B$. This implies that 
\begin{align}
&\big (\frac{v^j_{AB}(\Lambda, L)}{V(\Lambda, L)}\big )^{\frac{\Omega}{4} -1} < 1\\
\Rightarrow & \sum_{j=1}^{V(\Lambda, L)}\big (\frac{v^j_{AB}(\Lambda, L)}{V(\Lambda, L)}\big )^{\frac{\Omega}{4} -1} < V(\Lambda, L) \label{ppsh}
\end{align}
This suggests that we write the fine-tuning parameter in the following form
\begin{align}
F &= 1 - \prod_{A, B} \frac{\sum_{j=1}^{V(\Lambda, L)} \big (v^j_{AB}(\Lambda, L)/V(\Lambda, L)\big )^{\frac{\Omega}{4} -1}}{V(\Lambda, L)}\\
&= 1  - \prod_{A, B} \frac{\sum_{j=1}^{V(\Lambda, L)} \big (v^j_{AB}(\Lambda, L)/V(\Lambda, L)\big )^{N^22^{2N-2} -1}}{V(\Lambda, L)} \label{result}
\end{align}
For $N =1$, $F =0$ (there is no fine tuning). From equations (\ref{ppsh}) and (\ref{result}), we see that $0< F <1$ for any $N >1$. This implies that for finite $N$, unlike the superdeterministic model that satisfied (\ref{m}) and (\ref{n}), the model is not completely fine tuned (but still fine tuned). Lastly, $F = 1$ in the limit $N \to \infty$. This is because $\big (v^j_{AB}(\Lambda, L)/V(\Lambda, L)\big )^m \to 0$ as $m \to \infty$. Note that both $v^j_{AB}(\Lambda, L)$ and $V(\Lambda, L)$ do not depend on $N$, and that the result $F =1$ does not depend on the condition $\Lambda \to \infty$.\\

\section{Conspiracy as spontaneous entropy drop} \label{alu3}
In the previous section, we quantified the conspiratorial character of superdeterministic models of our scenario in terms of a fine tuning paramter $F$. The definition of $F$ rests on the concept of quantum nonequilibrium since $V(\Lambda)$ includes contributions from \textit{all} possible configurations of the hidden-variable distribution. For example, consider two distributions $p(\lambda|\{\alpha\}', \{\beta\}', \gamma_A', \gamma_B')$ and $p(\lambda|\{\alpha\}'', \{\beta\}'', \gamma_A'', \gamma_B'')$, where $M_A(\{\alpha\}', \gamma_A') = M_A(\{\alpha\}'', \gamma_A'')$ and $M_B(\{\beta\}', \gamma_B') = M_B(\{\beta\}'', \gamma_B'')$. For most of the configurations of $p(\lambda|\{\alpha\}', \{\beta\}', \gamma_A', \gamma_B')$ and $p(\lambda|\{\alpha\}'', \{\beta\}'', \gamma_A'', \gamma_B'')$, equation (\ref{i2}) will not be satisfied. But the definition of $F$ includes all these configurations as well in the quantity $V(\Lambda)^{\Omega}$. This is natural in a nonequilibrium framework, where all hidden-variable distributions are possible in principle. In this section, we give a quantification of superdeterministic conspiracy without using nonequilibrium distributions\footnote{Although there are compelling theoretical reasons \cite{genon, valentinu, valentiniastro} that motivate nonequilibrium extensions of hidden-variable models, a debate about nonequilibrium, for the purposes of this article, is secondary to proving that superdeterministic models are unequivocally conspiratorial.}. This approach is more general also in the sense that $\lambda$ and $p(\lambda|\{\alpha\}, \{\beta\}, \gamma_A, \gamma_B)$ do not have to be discrete. On the other hand, this approach is more limited in that it is applicable only to superdeterministic models that satisfy the constraint (\ref{m}). This approach is, however, still widely applicable as all the superdeterministic models that have been proposed in the literature so far posit a single distribution $p(\lambda| M_A, M_B)$. Therefore, those models satisfy both constraints (\ref{m}) and (\ref{n}) when extended to our scenario.\\

To begin, we first remind ourselves of the conspiratorial nature of apparent-signalling in superdeterminism, as discussed in $\mathcal{A}$. The violation of marginal-independence (or formal no-signalling) does not imply actual signalling in a superdeterministic context. The local marginals and the distant settings are correlated, but the distant setting does not causally affect the local marginal. The correlation is set up by the initial conditions. This appears to be a signal, because the distant setting can be controlled by an experimenter and the local marginal is correlated with the experimenter's decisions. We, therefore, term superdeterministic signalling as apparent signalling. We will now argue that an analogous situation occurs in superdeterministic models even in equilibrium. \\

Consider a superdeterministic model of our scenario that satisfies (\ref{m}). Equation (\ref{m}) implies that there are runs for which $\lambda$ is correlated only with the outputs $\alpha_i$ and $\beta_j$ (for a given pair $(i, j)$), and further, that these runs occur only when the experimenters choose $\gamma_A =i $ and $\gamma_B = j$. We may say that these runs belong to a \textit{sub-ensemble} $E = (i, j)$ having the hidden-variables distribution $\rho(\lambda| \alpha_{i}, \beta_{j}, \gamma_A = i, \gamma_B = j)$. There will be $N^2$ such sub-ensembles. However, although a run belonging to $(i, j)$ only occurs when the experimenters choose $\gamma_A =i $ and $\gamma_B = j$, there is no causal relationship between $\gamma_A$, $\gamma_B$ and the sub-ensemble $E$. Therefore, as in the case of apparent signalling, although it appears that the experimenters' choices determine which sub-ensemble a particular run belongs to, it is simply the case that the initial conditions have arranged a functional relationship $E = (\gamma_A, \gamma_B)$.\\

Note that this is true regardless of whether $\gamma_A$, $\gamma_B$ are determined by human experimenters, computer algorithms or any other device. It is also independent of $N$. Therefore, we have a situation where each setting mechanism is \textit{coincidentally} used only when it is correlated with the hidden variables. This appears conspiratorial intuitively. Below, we quantify this intuition as a spontaneous decrease of an appropriately-defined subjective `entropy' at the hidden-variables level.\\

Consider a superdeterministic model of our scenario that satisfies (\ref{m}). Let there be $N_0$ runs in total. Each run belongs to a particular sub-ensemble $E =(i, j)$ out of $N^2$ possibilities. The number of possible sequences of sub-ensembles over the $N_0$ runs is 
\begin{align}
W = N^{2N_0}
\end{align}
In quantum equilibrium, the experimenters cannot know which sequence, out of the $W$ possibilities, is actually the case. Suppose that by some means the experimenters find out the exact sequence. Then, they can choose $\gamma_A$ and $\gamma_B$ values that do not match this sequence. In that case, the Bell correlations will not be reproduced. But the measurement statistics cannot violate quantum predictions if quantum equilibrium is assumed. Therefore, quantum equilibrium precludes knowledge of the sub-ensemble sequence by the experimenters. The experimenters may assign subjective (Bayesian) probabilities to each possible sequence of sub-ensembles. Say the sequences are labelled by $k$, where $k \in \{1,2,...W\}$. Then these subjective probabilities $p(k)$ may be associated with a formal mathematical entropy
\begin{align}
S = -\sum_{k=1}^W p(k) \log_2 p(k) \label{bh}
\end{align} 

This is the Shannon entropy of the distribution of sub-ensemble sequences over the $N_0$ runs. Note that this entropy is subjective in the sense that there is only one actual sequence for the experiment, but the experimenters have assigned probabilities $p(k)$ to different sequences out of ignorance of the actual sequence.\\

\begin{figure}
\includegraphics[scale=0.3]{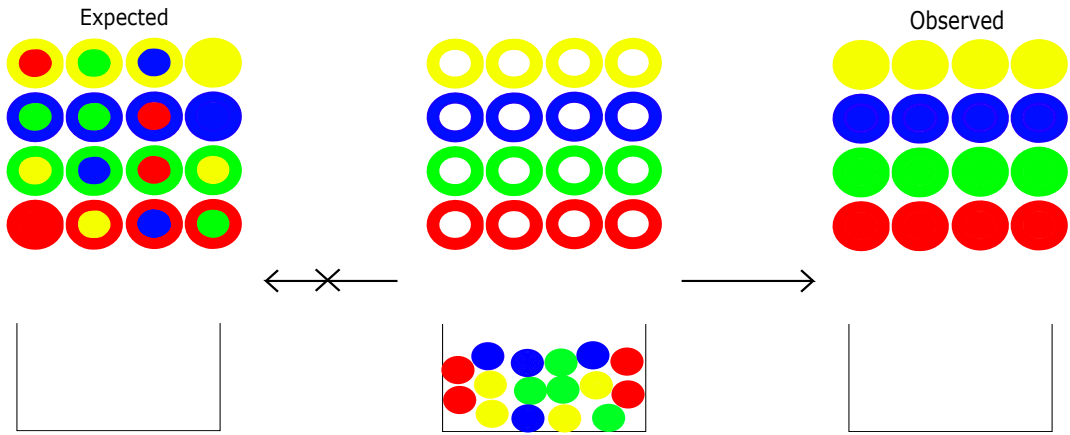}
\caption{Illustration of spontaneous entropy drop in superdeterministic models. Each solid ball represents a particular pair $(\gamma_A, \gamma_B)$ of setting mechanisms. Its colour represents the value of $(\gamma_A, \gamma_B)$. Each socket represents an experimental run and its colour represents the sub-ensemble $E =(i, j)$ for that run. Initially, as shown in the middle figure, there are a number of balls kept in a b, arranged in no particular order, along with an equal number of sockets. The superdeterministic model is consistent only if $E = (\gamma_A, \gamma_B)$. Pictorially, this corresponds to matching the colours of both the balls and the sockets. However, the experimenters do not have access to the information `which run belongs to which sub-ensemble' in quantum equilibrium. That is, the experimenters know the colours of the balls but not the colours of the sockets. One would expect that a situation like that shown in the left-hand figure would therefore arise. However the experimenters always find out that they have successfully matched each and every ball with its corresponding socket, as shown in the right-hand figure. This is because the initial conditions ensure $E = (\gamma_A, \gamma_B)$ by a sequence of exact coincidences according to superdeterminism. We quantify the amount of correlation required for such a superdeterministic mechanism as a drop of an appropriately defined formal entropy.}
\end{figure}

Next, let the experimenters determine the setting mechanisms ($\gamma_A$ and $\gamma_B$) for each run. Constraint (\ref{m}) implies that $E = (\gamma_A, \gamma_B)$ for each run. Therefore, one of the probabilities $p(k)$ then equals 1 and the rest are all 0. The associated entropy $S = -1\log_2 1 = 0$, and the entropy has therefore decreased by
\begin{align}
\Delta S =\sum_{k=1}^{W} p(k) \log_2 p(k)
\end{align}

As the experimenter at wing A (B) has to make no conscious effort to match $\gamma_A$ ($\gamma_B$) values with the sub-ensemble sequence, it is fair to say that this entropy decrease is \textit{spontaneous} (see Fig. 4). Note that $\Delta S = - H(k : \{\gamma_A, \gamma_B\})$, where $H(k : \{\gamma_A, \gamma_B\})$ is the mutual information \cite{nielsen} between the variables $k$ and $\{\gamma_A, \gamma_B\}$ (the set of $\gamma_A, \gamma_B$ values for $N_0$ runs). Note also that $\Delta S$ does not depend on $p(\lambda|M_A, M_B)$ or $p(\{\gamma_A, \gamma_B\})$. \\

It is important to assume equation (\ref{m}) to derive this entropy decrease. Suppose that equation (\ref{m}) is not true. There are two possibilities in this case. First, $\lambda$ is correlated to multiple setting mechanisms. In that case, the sub-ensembles and the associated entropy $S$ cannot be defined. Second, $\lambda$ is correlated to individual setting mechanisms, but these are not the ones chosen by the experimenters. In that case, the experimenters can assign probabilities $p(k)$ to different sub-ensemble sequences as before. But, after choosing $\gamma_A$ and $\gamma_B$ for each run, the entropy $S \neq 0$. \\

We now evaluate the net spontaneous entropy drop for the simplest case that $p(k)=1/W \textbf{ } \forall \textit{ } k$. In this case, the net entropy drop can be thought of as the sum over $N_0$ runs of the individual entropy drop for each run. The net entropy drop is found to be 
\begin{align}
\Delta S = -2N_0\log_2 N \label{alas}
\end{align}

In terms of mutual information, equation (\ref{alas}) implies that $H(E : \gamma_A, \gamma_B) = 2\log_2 N$. Let us compare this with the mutual information between $\lambda$ and the measurement settings for superdeterministic models of a Bell scenario for $N=1$. Ref. \cite{hall20} reports a superdeterministic model which needs only $ \sim 0.08$ bits of mutual information between $\lambda$ and the measurement settings. On the other hand, equation (\ref{alas}) implies that for $N = 16$, $H(E : \gamma_A, \gamma_B)  = 8$ bits, which is $\sim 100$ times the mutual information required between $\lambda$ and the measurement settings for the aforementioned superdeterministic model. The extremely strong correlation between $E$ and $\gamma_A, \gamma_B$ arranged by the initial conditions is a conspiratorial feature of superdeterministic models of our scenario. In the limit $N \to \infty$, we have $\Delta S \to \infty$ and in this sense we may say that the superdeterministic model becomes arbitrarily conspiratorial. \\

As in the previous approach, nonlocal and retrocausal models are not conspiratorial according to this approach. This is because the variables $\{\alpha\}, \gamma_A$ and $\{\beta\}, \gamma_B$ can be excluded from these models (see section \ref{alu1}), so that there are no sub-ensembles associated with different setting mechanisms. \\

\section{Discussion} \label{alu4}
We have provided two different ways to quantify the often-alleged conspiratorial character of superdeterministic models, using intuitive suggestions from a previous article $\mathcal{A}$. The first approach utilises the idea that different setting mechanisms lead to different measurement statistics in general for a nonequilibrium superdeterministic model. We have shown that superdeterministic models must be fine-tuned so that the measurement statistics depend on the measurement settings, but not on how these settings are chosen. No features of quantum mechanics, such as the Born rule, marginal-independence (formal no-signalling), non-contextuality etc. have been assumed in deriving this result. Further, we have quantified the fine tuning by defining an overhead fine-tuning parameter $F$. We calculated $F$ for two different kinds of superdeterministic models of our scenario. We first considered superdeterministic models as usually proposed in the literature, with a single hidden-variables distribution $p(\lambda|M_A, M_B)$. The fine tuning for such a model is drastic as $F = 1$ (corresponding to `completely fine tuned') for any $N > 1$. This result makes use of the condition $\Lambda \to \infty$ from Hardy's `excess-baggage' theorem \cite{hardycess}. We have also considered more general superdeterministic models of our scenario. For these general models, $F < 1$ for any $N > 1$. In the limit $N \to \infty$, $F = 1$. This result does not depend on the condition $\Lambda \to \infty$.\\

It is useful to discuss some possible misunderstandings about this result. One might argue that, in a nonequilibrium framework, an enormous amount of fine-tuning is inevitable anyway just to restrict outcomes to the Born rule. A bit more fine-tuning is therefore perhaps nothing to be concerned about. A counter example to such an argument would be the pilot-wave theory, where no fine tuning is necessary to reproduce the Born rule, even when nonequilibrium is allowed. This is because a nonequilibrium distribution in pilot-wave theory dynamically relaxes to the equilibrium distribution over time, given appropriate assumptions about the initial distribution \cite{valentinI, royalvale}. Note that equilibration as an explanation of fine tuning is noted in the causal modelling literature as well \cite{finedash, woodspek}. Another possibility is that one might argue that an un-fine-tuned $(F=0)$ classical universe will have a high-entropy past. The fine-tuning argument applied to our universe would therefore prove that a low-entropy past is very unlikely, disproving the second law of thermodynamics. However, there remain ambiguities about whether the entropy of the whole universe is a well-defined concept because of the role of gravity \cite{earman}. In the early universe, this problem is severe because gravitational effects cannot be ignored. Therefore, it is not clear at present whether an explanation of thermodynamics in terms of a low-entropy past of the universe can be given.\\

The parameter $F$ quantifies how special the hidden-variables distribution has to be so that the measurement statistics depend only on the measurement settings. Clearly, the notion of `special' is meaningful only when there are multiple possible distributions, that is, if nonequilibrium is allowed (at least in principle). Those who consider nonequilibrium distributions to be unmotivated may find this approach to be artificial. We have, therefore, provided a second way to quantify the conspiracy, which does not use nonequilibrium distributions. \\

The second approach uses another idea from $\mathcal{A}$ that the violation of marginal independence (formal no-signalling) in superdeterministic models in nonequilibrium does not imply actual signalling. This leads to the intuitively conspiratorial situation where two experimenters can use marginal dependence as a practical signalling procedure without actually sending signals. In this case, the entire sequence of messages exchanged between the experimenters is interpreted as a statistical coincidence. Analogous to this, for a superdeterministic model of our scenario restricted to equilibrium, there occurs a one-to-one correspondence between the experimenters' choice of setting mechanisms and the sub-ensemble values for all runs. But this correspondence is, again, only a sequence of statistical coincidences. This appears conspiratorial as it shows that the initial conditions in superdeterministic models of our scenario need to arrange extremely strong correlations. Note that there have been several studies of the amount of correlation required between $\lambda$ and the measurement settings in superdeterministic models \cite{hallrelax, hall16, hall19, hall20}. A low correlation value might be interpreted in favour of superdeterministic models on grounds of efficiency (as has been suggested for retrocausal models in ref. \cite{hall20}). Our argument shows that this would be incorrect: there has to be an extremely strong correlation (one-to-one correspondence) between the experimenters' choices and the sub-ensemble values in our scenario regardless of the strength of correlation between $\lambda$ and the measurement settings. We quantify the correlation as a formal entropy drop $\Delta S$ defined at the hidden-variables level. We show that $\Delta S \to \infty$ logarithmically fast with $N$, and in this sense we may say that the equilibrium superdeterministic model becomes arbitrarily conspiratorial as $N \to \infty$. \\

Nonlocal and retrocausal models of our scenario are non-conspiratorial by definition according to either approach, as the variables related to the different setting mechanisms and experimenters' choices can be eliminated from the model description. Note that nonlocal, retrocausal and superdeterministic models are all conspiratorial in the sense that they require fine-tuning in order to satisfy marginal independence (formal no-signalling) \cite{genon, woodspek}, unless a dynamical explanation of finetuning is given \cite{valentinI, royalvale, teenv}. Our results here are concerned only with the notion of superdeterministic conspiracy (see $\mathcal{A}$).\\

We have quantitatively proven that superdeterministic models are conspiratorial in two different ways. What are the possible options for a future superdeterminist? Let us consider the possible options for circumventing our two arguments separately. Consider first the fine-tuning argument. One option to evade fine tuning might be to eschew the possibility of quantum nonequilibrium. This will involve abandoning the distinction between initial conditions and laws, which is a central principle in scientific theories. Palmer seems to advocate such an approach \cite{palmer09, palmerend}. However, his model still appears to assume an equilibrium distribution over certain parameters, a distribution which could in principle be different\footnote{See ref. \cite{sinp} for a detailed discussion of Palmer's proposal in a hidden-variables framework.}. Another option might be to build a superdeterministic model with a dynamical relaxation mechanism such that an arbitrary distribution that does not satisfy equation (\ref{ii}) evolves over time to a distribution that satisfies it.\footnote{Given appropriate assumptions about the initial distribution, as in classical statistical mechanics \cite{tallman}.} For example, dynamical relaxation to the equilibrium distribution is a natural consequence of the evolution law in pilot-wave theory \cite{valentinI, royalvale, teenv}. In such a superdeterministic model, however, small deviations from equilibrium may occur. This is because relaxation to the equilibrium distribution may get delayed due to various factors \cite{valentiniastro, teenv}. For quantum systems trapped in nonequilibrium, the superdeterministic model will predict the violation of condition (\ref{ii}). One may use this to experimentally test such a superdeterministic model in the nonequilibrium regime. Let us now consider the spontaneous entropy drop argument. This approach does not depend on quantum nonequilibrium, but it is applicable only to models that satisfy condition (\ref{m}). Therefore, a superdeterministic model that violates condition (\ref{m}) can circumvent this argument. However, in such a model $\lambda$ will not only be correlated with the physical variables that determine the measurement settings, but also with other physical variables that might have been used by the experimenters to choose the settings. These other variables could in principle be any variable from the entire observable universe. A superdeterministic model that violates (\ref{m}) would therefore appear to exchange one kind of conspiracy for another. \\

We conclude that, at least as they are currently understood, superdeterministic models are conspiratorial. On this basis it seems fair to conclude that superdeterminism, as a possible explanation of the Bell correlations, is scientifically unattractive.

\acknowledgements
We are grateful to the anonymous referees for helpful comments.

\bibliographystyle{bhak}
\bibliography{bib}

\begin{thebibliography}{10}

\bibitem{1st}
I.~Sen and A.~Valentini.
\newblock {Superdeterministic hidden-variables models I: nonequilibrium and
  signalling}.
\newblock {\em arXiv:2003.11989}, 2020.

\bibitem{valentiniastro}
A.~Valentini.
\newblock {Astrophysical and cosmological tests of quantum theory}.
\newblock {\em {J. Phys. A}}, {40}({12}), {2007}.

\bibitem{teenv}
A.~Valentini.
\newblock {Foundations of statistical mechanics and the status of the Born rule
  in de Broglie-Bohm pilot-wave theory}.
\newblock In {\em Statistical Mechanics and Scientific Explanation:
  Determinism, Indeterminism and Laws of Nature}. World Scientific, 2020.

\bibitem{genon}
A.~Valentini.
\newblock {Signal-locality in hidden-variables theories}.
\newblock {\em {Phys. Lett. A}}, {297}({5-6}), {2002}.

\bibitem{woodspek}
C.~J. Wood and R.~W. Spekkens.
\newblock {The lesson of causal discovery algorithms for quantum correlations:
  Causal explanations of Bell-inequality violations require fine-tuning}.
\newblock {\em {New J. Phys.}}, {17}({3}), {2015}.

\bibitem{bell}
J.~S. Bell.
\newblock {\em {Speakable and unspeakable in quantum mechanics: Collected
  papers on quantum philosophy}}.
\newblock {Cambridge Univ. Press}, {2004}.

\bibitem{jearl}
J.~Pearl.
\newblock {\em Causality}.
\newblock Cambridge university press, 2009.

\bibitem{valentinI}
A.~Valentini.
\newblock {Signal-locality, uncertainty, and the subquantum H-theorem. I}.
\newblock {\em {Phys. Lett. A}}, {156}({1-2}), {1991}.

\bibitem{royalvale}
A.~Valentini and H.~Westman.
\newblock {Dynamical origin of quantum probabilities}.
\newblock {\em Proc. R. Soc. A}, 461(2053):253--272, 2005.

\bibitem{hardycess}
L.~Hardy.
\newblock {Quantum ontological excess baggage}.
\newblock {\em {Stud. Hist. Phil. Sci. B}}, 35, 2004.

\bibitem{valentinu}
A.~Valentini.
\newblock {Universal signature of non-quantum systems}.
\newblock {\em {Phys. Lett. A}}, {332}({3}), {2004}.

\bibitem{nielsen}
M.~A. Nielsen and I.~L. Chuang.
\newblock {\em {Quantum computation and quantum information}}.
\newblock Cambridge Univ. Press, 2000.

\bibitem{finedash}
D.~Dash.
\newblock {Restructuring Dynamic Causal Systems in Equilibrium}.
\newblock In {\em AISTATS}. Citeseer, 2005.

\bibitem{earman}
J.~Earman.
\newblock {The “past hypothesis”: Not even false}.
\newblock {\em Stud. Hist. Philos. Sci. B}, 37(3):399--430, 2006.

\bibitem{hallrelax}
M.~J. Hall.
\newblock {Relaxed bell inequalities and kochen-specker theorems}.
\newblock {\em Phys. Rev. A}, 84(2):022102, 2011.

\bibitem{hall16}
M.~J. Hall.
\newblock {The significance of measurement independence for Bell inequalities
  and locality}.
\newblock In {\em {At the Frontier of Spacetime}}. {Springer}, {2016}.

\bibitem{hall19}
A.~S. Friedman, A.~H. Guth, M.~J. Hall, D.~I. Kaiser, and J.~Gallicchio.
\newblock {Relaxed Bell inequalities with arbitrary measurement dependence for
  each observer}.
\newblock {\em Phys. Rev. A}, 99(1):012121, 2019.

\bibitem{hall20}
M.~J. Hall and C.~Branciard.
\newblock {Measurement-dependence cost for Bell nonlocality: causal vs
  retrocausal models}.
\newblock {\em arXiv: 2007.11903}, 2020.

\bibitem{palmer09}
T.~Palmer.
\newblock {The Invariant Set Postulate: a new geometric framework for the
  foundations of quantum theory and the role played by gravity}.
\newblock {\em Proc. R. Soc. A}, 465(2110):3165--3185, 2009.

\bibitem{palmerend}
T.~Palmer.
\newblock {Discretization of the Bloch sphere, fractal invariant sets and
  Bell’s theorem}.
\newblock {\em Proc. R. Soc. A}, 476(2236):20190350, 2020.

\bibitem{sinp}
I.~Sen.
\newblock {Discussion of Palmer's superdeterministic proposal in a
  hidden-variables setting}.
\newblock In preparation.

\bibitem{tallman}
R.~C. Tolman.
\newblock {\em {The principles of statistical mechanics}}.
\newblock Oxford Univ. Press, 1938.

\end{thebibliography}

\end{document}